\documentclass[aps,prl,graphics,twocolumn,groupedaddress]{revtex4}

\usepackage{graphicx}
\usepackage{dcolumn}
\usepackage{bm}
\usepackage{epsf}

\usepackage{graphics}

\begin{document}

\title{Surface tension of liquids near the critical point }

\author{ N.V. Brilliantov$^{1,2}$ and J.M. Rubi$^{1}$}

\affiliation{ 
  $^1$Departament de Fisika Fonamental, Universitat de Barcelona, 
  Av.Diagonal
  647, 08028 Barcelona, Spain} 
\affiliation{$^2$Moscow State University, Physics Department,
Moscow 119899, Russia}
\date{\today}

\begin{abstract}
A simple analytical microscopic expression for the surface tension of
liquids $\gamma$ is obtained which is in a good agreement with 
available data of numerical experiments. We apply the integral 
transformation that   maps the fluid Hamiltonian onto the 
field-theoretical Hamiltonian and show that the order 
parameter of the effective Hamiltonian corresponds to the one-body 
potential in the fluid. Revealing the physical meaning of the order 
parameter allows calculation  of $\gamma$.

\end{abstract}


\maketitle

The modern theory of nonuniform fluids is based mainly on the integral 
equations 
hierarchy or density functional theory \cite{Bookinhom}. Recent 
applications of the latter theory to the problems of interface formation 
and surface tension demonstrated its accuracy and effectiveness 
\cite{Evans:92,Forstman:97}. Although the density functional theory often 
uses {\it ad hoc} approximations, rather than regular expansion schemes 
\cite{Evans:92}, it is essentially the first-principle 
theory: No model parameter are involved, and for a given interparticle 
potential, temperature, etc., the density distribution and surface tension 
may be obtained  \cite{Evans:92,Forstman:97}. Being physically very clear 
and straightforward,  the density functional theory is not very easy in 
practical applications, since sophisticated numerical analysis is required 
\cite{Evans:92,Forstman:97}. It does not generally provide any analytical 
expression for the surface tension. The other limitation of the density 
functional theory  refers to its strictly mean-field nature  \cite{Evans:92}. 
Apart from its phenomenological version \cite{FiskWidom:69} (where model 
parameters were used) the density functional theory is not suited to study 
the critical region, where the critical fluctuations play a clue role.

In somewhat alternative,  field-theoretical approach \cite{Amit:78}, an 
order parameter $\phi({\bf r})$ is introduced and functional 
dependence of the free energy (or of the effective Hamiltonian) on 
$\phi({\bf r})$ is postulated. For the most frequently used square-gradient 
model it reads (e.g. \cite{Amit:78,BrezinFeng:84})

\begin{equation}
\label{betaH}
\beta {\cal H} = \int d{\bf r} \left[(\kappa/2) (\nabla \phi)^2  + 
V(\phi) \right]
\end{equation}
where the first term accounts for the free energy penalty due to inhomogeneity 
and $V(\phi)$ is chosen to 
mimic a possibility of the two-phase coexistence. Its simplest choice 
corresponds to the Landau-Ginzburg-Wilson (LGW) Hamiltonian:
\begin{equation} 
\label{LGWdef}
V(\phi)=(a_2/2!) \phi^2({\bf r}) + (u_4/4!) \phi^4 ({\bf r})
- h ({\bf r}) \phi ({\bf r}) \, , 
\end{equation} 
with the free energy, 
$-\beta F = \log \int {\cal D} \phi \exp(-\beta {\cal H})$ ($\beta =1/k_BT$), 
which accounts for all possible distributions of the order parameter, and not 
only for the extremal one, as in the density functional theory. 
 The powerful field-theoretical methods, elaborated especially 
for the LGW Hamiltonian \cite{Amit:78,BrezinFeng:84} may be then applied to 
describe the critical fluctuations, and  a simple analytical 
expression for the surface tension may be derived  
\cite{BrezinFeng:84,Igna:2000}.

The order parameter $\phi ({\bf r})$ in (\ref{betaH},\ref{LGWdef}) 
 may be of any physical nature, (density, composition of the 
fluid  mixture \cite{Igna:2000},  magnetization  \cite{BrezinFeng:84}, etc.), 
while the coefficients of the Hamiltonian ${\cal H}$
are some model parameters for which microscopic expressions are not provided. 
Therefore  the field theoretical approach is not now, strictly speaking, a 
microscopic theory. 

The present study is addressed to overcome this flaw. 
We obtain an effective field theoretical Hamiltonian for the fluid with 
microscopic expressions for its coefficients. 
We show that the order parameter of this Hamiltonian corresponds to the 
one-body  microscopic potential, and its average to the first-order direct 
correlation function. Revealing the physical nature of the order parameter 
we obtain an analytical expression for the surface tension which is in a good 
agreement with numerical experiments.

Referring for detail to \cite{Brill:98,BrilRubi:2001} we briefly sketch 
derivation of the effective LGW Hamiltonian, which is similar to that of 
Hubbard and Schofield \cite{hubbard}. We start from the fluid Hamiltonian 
$H=H_R+H_A+H_{\rm ex}$:

\begin{equation}
H=\sum_{i<j} v_r ({\bf r}_{ij}) - \sum_{i<j} v({\bf r}_{ij}) 
+\sum_j g({\bf r}_j)
\label{Ham}
\end{equation}
where $v_r(r)$ denotes the  repulsive part of the  
interaction potential, $-v(r)$ denotes the attractive part and $g({\bf r})$ 
denotes the external potential;   
$\{ {\bf r}_j \}$ are coordinates of the particles and 
${\bf r}_{ij}={\bf r}_i-{\bf r}_j$. 
The last two terms of the fluid Hamiltonian (\ref{Ham})  may be written 
using the Fourier transforms of the density fluctuations, 
$n_{\bf k}=(1/\sqrt{\Omega}) \sum_{j=1}^{N} e^{-i {\bf k} \cdot {\bf r}_j}$, 
of the attractive potential, 
$v_k=\int v(r)e^{-i {\bf k} \cdot {\bf r}} d {\bf r} $ and of the 
external potential 
$g_k=\int g({\bf r})e^{-i {\bf k} \cdot {\bf r}} d {\bf r} $ as 

\begin{equation}
-(1/2) \sum_{\bf k } v_k n_{\bf k} n_{-{\bf k}}+(1/2)v(0)N 
+ \sum_{\bf k } g_{\bf k} n_{-{\bf k}}\, ,
\end{equation}
where $N$ is the number of particles,  $\Omega=L^3$ is the volume of the 
system ($L \to \infty$), and summation over $k_l=\frac{2\pi}{L}n_l$ 
with  $l=x,y,z$, and $n_l=0,\pm 1, \ldots$ is implied.
Let $\mu$ be the chemical potential of the 
system with the ``complete'' Hamiltonian, (\ref{Ham}), and $\mu_R$ be the 
chemical potential of the ``reference'' system, with the Hamiltonian, 
$H_R$, which has only repulsive interactions. Then the grand partition 
function, $\Xi$,  may be expressed in terms of the grand partition 
function, $\Xi_R$ of the reference fluid as ~\cite{hubbard}

\begin{equation}
\Xi=\Xi_R \left\langle \exp \left\{ \beta \mu' N+\beta 
\sum_{\bf k }\left[ 
\frac{v_k}{2} n_{\bf k} n_{-{\bf k }} - 
n_{\bf k} g_{-{\bf k}} \right] \right\} \right\rangle_R , \nonumber 
\label{Xi}
\end{equation}
where $\mu'=\mu-\mu_R+\frac12 v(0)$ and   
$\left\langle  \, \, \, \right\rangle_R$ denotes an  average over the 
reference system at temperature $T$ with chemical potential $\mu_R$. 
Using the identity:  
$\exp(\frac{1}{2} a^2x^2-bx)=(2 \pi a^2)^{-1/2} 
\int_{-\infty}^{+\infty} \exp [-\frac{1}{2} (y+b)^2/a^2 + xy]dy$, 
we  obtain for the ratio $Q=\Xi/\Xi_R$:

\begin{eqnarray}
\label{Q2}
&&Q   \propto \int  
\prod_{\bf k }  d\phi_{\bf k }
\left\langle 
 \exp \left\{ \sum_{\bf k}
 \phi_{\bf k } n_{-{\bf k} } 
\right\} \right\rangle_R 
\exp \left\{ \, \frac{\mu'}{v_0} \Omega^{1/2}\phi_0 \right. \nonumber \\ 
&& \left. -\frac12 ~\beta^{-1} \sum_{\bf k} 
v_k^{-1}
\left( \phi_{\bf k } + \beta g_{\bf k } \right)
\left( \phi_{-{\bf k} } + \beta g_{-{\bf k} } \right) \right\}\, ;
\end{eqnarray}
here integration is to be performed under the constrain 
$\phi_{-{\bf k} }=\phi_{\bf k }^*$,  
and a factor which does not  affect the subsequent 
analysis is omitted. Applying the cumulant  
theorem to the factor 
$\left\langle \exp \left\{\sum_{\bf k } 
\phi_{\bf k }n_{-{\bf k }}\right\}\right\rangle _R$ 
one can write \cite{hubbard}

\begin{eqnarray}
\label{Q3}
&& Q   \propto \int  \prod_{\bf k }  d\phi_{\bf k } \exp(- \beta {\cal H} ) \\
&&\beta {\cal H}= -\tilde{h} \, \Omega^{1/2} \phi_0 
+ \sum_{n=2}^{\infty} \Omega^{1-n/2}
 \sum_{{\bf k}_1, \ldots {\bf k}_n} \tilde{u}_n \, 
\phi_{{\bf k}_1} \cdots \phi_{{\bf k}_n} \, , \nonumber 
\end{eqnarray}
where the coefficients of the effective field theoretical Hamiltonian 
${\cal H}$ read for $g({\bf r })=0$: 
 $\tilde{h}= \mu^{\prime} v_0^{-1} +\rho $, and 
\begin{eqnarray}
\label{defhun}
&&\tilde{u}_{2}({\bf k}_1,{\bf k}_2)
=(1/2!)  \, \delta_{{\bf k}_1 +{\bf k}_2,0} \left\{\beta^{-1}v_{k_1}^{-1}-
\left\langle n_{{\bf k}_1} n_{-{\bf k}_1} \right\rangle _{cR} \right\}\, , 
\nonumber \\
&&\tilde{u}_{n}\left({\bf k}_1, \ldots {\bf k}_n \right)=
-(\Omega^{n/2-1}/n!)  
\left\langle n_{{\bf k}_1} \cdots n_{{\bf k}_n} \right\rangle_{cR} 
\end{eqnarray}
for $n \ge 3$. 
Here $\left\langle \, \, \, \right\rangle _{cR}$ denotes the {\it cumulant} 
average calculated in the (homogeneous) reference system and 
$\rho=\Omega^{-1/2} < n_0 >_{cR}=N/\Omega$ is the fluid density. 
According to (\ref{Q3}), $Q$ has the form of a partition function. 

As it follows from (\ref{defhun}) the coefficients of ${\cal H}$ depend 
on correlation function of the reference fluid having only repulsive 
interactions. Using definitions of the particle correlation functions of 
fluids ~\cite{gray} one can express the cumulant averages 
$\left\langle n_{{\bf k}_1} \cdots n_{{\bf k}_n} \right\rangle _{cR}$, and 
thus the coefficients $\tilde{u}_{n}\left({\bf k}_1, \ldots {\bf k}_n \right)$
in terms of the Fourier transforms of the {\it connected} correlation 
functions $h_1,h_2, \ldots h_n$ of the reference system. These are defined as 
$h_1({\bf r}_1) \equiv \delta({\bf r}_1)$, 
$h_2({\bf r}_1,{\bf r}_2) \equiv g_2({\bf r}_1,{\bf r}_2)-1$, 
$h_3({\bf r}_1,{\bf r}_2,{\bf r}_3)= g_3({\bf r}_1,{\bf r}_2,{\bf r}_3)-
g_2({\bf r}_1,{\bf r}_2)-g_2({\bf r}_1,{\bf r}_3)-g_2({\bf r}_2,{\bf r}_3)+2$, 
etc., where $g_l({\bf r}_1, \ldots {\bf r}_l)$  are $l$-particle correlation 
functions ~\cite{gray}. In particular, the first few coefficients read 
\cite{Brill:98}

\begin{eqnarray}
\label{coef1}
&&\tilde{u}_{2} = (1/2!)\left[(\beta v_0)^{-1}-
\rho-\rho^2 \tilde{h}_2({\bf k}_1)
\right] \delta_{{\bf k}_1+{\bf k}_2,0}  \\
\label{coef2}
&&\tilde{u}_{3}=-(1/3!)  
\left\{\rho + \rho^2 \left[ \tilde{h}_2({\bf k}_1)+ 
\tilde{h}_2({\bf k}_2)+\tilde{h}_2({\bf k}_3) \right] \right.\nonumber \\
&&\left.+\rho^3 \tilde{h}_3({\bf k}_1;{\bf k}_2) \right\}
\delta_{{\bf k}_1+{\bf k}_2+{\bf k}_3,0} \, ,
\end{eqnarray}
where $\tilde{h}_l$ are the Fourier transforms of $h_l$. 

To obtain the conventional square gradient form of the effective Hamiltonian
we perform the small-$k$ expansion of the coefficients 
$\tilde{u}_{n}\left({\bf k}_1, \ldots {\bf k}_n \right)$. From the structure 
of the LGW Hamiltonian (\ref{betaH}), which has the only gradient term 
$(\nabla \phi)^2 \sim k^2 \phi_{\bf k } \phi_{-{\bf k} }$,  follows that 
only $\tilde{u}_2$ should be expanded as 
$\tilde{u}_2 =\tilde{u}_2(0)- \tilde{u}_2(0)^{\prime\prime}k^2 +\cdots$, while 
the other coefficients $\tilde{u}_n$, $n \ge 3$ should be taken at zero 
wave-vectors, as $\tilde{u}_{n}\left(0, 0, \ldots 0 \right)$ \cite{Brill:98}. 
Thus, as it is seen from (\ref{coef1},\ref{coef2}) only 
$\tilde{h}_2(0)$, $\tilde{h}_2(0)^{\prime\prime}$ and 
$\tilde{h}_l(0) \equiv \tilde{h}_l(0,0,\ldots 0)$, $l>2$, 
are needed. To find the latter  we use the chain relation
for the successive correlation functions \cite{Brill:98}

\begin{equation}
\label{relclus}
z_0 \, \rho  
\frac{ \partial}{\partial \rho} \rho^l \tilde{h}_l(0) = 
\rho^l \left[ l\, \tilde{h}_l(0) + \tilde{h}_{l+1} (0) \right] \, 
\end{equation}
expressing each $\tilde{h}_{l+1}(0)$ in terms of $\tilde{h}_{l}(0)$ and 
its density derivative. Here 
$z_0 \equiv \beta^{-1}\left( \partial P / \partial \rho \right)^{-1}_{\beta}$ 
is the reduced isothermal 
compressibility and $P$ is the pressure. With $\tilde{h}_1(0)=1$, 
 Eq.(\ref{relclus}) allows to express iteratively each of $\tilde{h}_l(0)$, 
(which refer  to the reference fluid) in terms of the 
reduced compressibility of the reference fluid $z_0$ and its density 
derivatives  
$\partial z_0/ \partial \rho$, $\partial^2 z_0/ \partial \rho^2$, etc.

For the reference system with the 
only repulsive interactions one can use the hard--sphere system with an 
appropriately chosen diameter ~\cite{gray}. For soft (not 
impulsive) repulsive forces a simple relation ~\cite{gray} 
$d=\int_0^{\sigma} \left[ 1- \exp \left( - \beta v_r(r)\right) \right]$
gives the effective diameter of the hard-sphere system, corresponding to a  
repulsive potential $v_r(r)$ vanishing at $r \ge \sigma$. 
The fairly accurate Carnahan-Starling equation of state for this 
system~\cite{gray} yields for the reduced compressibility 
\begin{equation}
\label{z0} 
z_0=\left(1-\eta \right)^4/ \left( 1+4\eta +4\eta^2 -4\eta^3 +\eta^4 \right)
\end{equation}
where $\eta=\frac{ \pi}{6} d^3 \rho$. For the hard-sphere reference system 
one can also find $\tilde{h}_2(0)^{\prime\prime}$. This may be done expressing 
$\tilde{h}_2(k)$ in terms of the direct correlation function 
$\tilde{c}_2(k)$, as 
$\tilde{h}_2(k)=\tilde{c}_2(k)/\left[1-\rho \tilde{c}_2(k) \right]$ 
~\cite{gray},  expanding $\tilde{c}_2(k)$ as 
$\tilde{c}_2(k)=\tilde{c}_2(0)-\tilde{c}_2(0)''k^2+\cdots$, and using 
$\tilde{c}_2(0)''=-(\pi d^5/120)
\left(16-11\eta +4 \eta^2 \right)\left( 1-\eta \right)^{-4}$ which may be  
obtained from the Wertheim-Thiele solution \cite{gray} for the direct 
correlation function \cite{Brill:98}. 

Following these lines we find explicite expressions for all coefficients of the
effective Hamiltonian (\ref{Q3}). To recast  this into conventional 
form (\ref{betaH},\ref{LGWdef}) we (i)~perform transformation from variables 
$\phi_{\bf k}$ to the space-dependent field $\phi({\bf r})$, (ii)~omit all 
terms in (\ref{Q3}) with powers of the field higher that fourth, and 
(iii)remove the cubic term with respect to the field, which $V(\phi)$ does not 
contain. This may be done by the shift $\phi \to \phi + \bar{\phi}$, with 
$\bar{\phi}$ chosen to make the cubic term vanish. As the result, 
we arrive at the effective 
Hamiltonian (\ref{betaH}) with {\it microscopic} expressions 
for its coefficients. The coefficients of $V(\phi)$  read

\begin{eqnarray}
\label{coefinal}
&&u_4=-\rho z_0 [z_1^2+z_0(z_0+4z_1+z_2)] \nonumber \\
&&a_2=(\beta v_0)^{-1}-\rho(z_0+z_3^2/2u_4) \\
&&h=\mu^{\prime} v_0^{-1} 
+\rho(a_2+z_3^2/6u_4)z_3/u_4 +\rho  \nonumber \, .
\end{eqnarray}
with  $z_0$ given by (\ref{z0}), 
$z_1 \equiv \rho \partial z_0/ \partial \rho$,
$z_2 \equiv \rho^2 \partial^2 z_0/ \partial \rho^2$ and 
$z_3 \equiv - \rho z_0 (z_0+z_1)$, while the coefficient $\kappa$ reads
 
\begin{equation}
\label{kappa}
\kappa=(3/40 \pi d) 
\left[ \lambda_{\rm eff}^2/\beta \epsilon_{\rm eff}-B \right]
\end{equation}
where 
$B=4 \eta^2(1-\eta)^4(16-11\eta+4 \eta^2)/(1+4\eta+4\eta^2-4\eta^3+\eta^4)^2$ 
and constants $\epsilon_{\rm eff}$ and $\lambda_{\rm eff}$ 
(which appear due to the expansion of $\tilde{u}_2$) characterize the 
effective  depth and effective width of the attractive  part $v(r)$:
$\epsilon_{\rm eff}=(4\pi d^3/3)^{-1}\int v(r) d{\bf r}$ and 
$\lambda_{\rm eff}^2=(3v_0d^2/5)^{-1}\int v(r) r^2 d{\bf r}$.

To this point, we have derived the effective field theoretical Hamiltonian 
with explicit microscopic expressions for its coefficients. Now we show that 
the order parameter $\phi({\bf r})$ which has been introduced as a 
``technical'' variable of the integral transformation has a clear physical 
meaning: it is related to the local microscopic potential. 

To show this we notice that from (\ref{Xi}) it directly follows 
$\partial \log \Xi / \partial g_{-{\bf k}}= 
\partial \log Q / \partial g_{-{\bf k}}=-\beta 
\left \langle n_{\bf k} \right \rangle$; on the other hand  
Eq.(\ref{Q2}) allows to write for $g \to 0$:
$\partial \log Q / \partial g_{-{\bf k}} = 
-v_k^{-1}\left \langle \phi_{\bf k} \right \rangle$, where the averaging is 
to be understood as the integration over all distributions of the order 
parameter. Recalling that we have shifted the field by the constant 
$\bar{\phi}$, we conclude that $\langle \phi_{\bf k} \rangle = 
\beta v_k \langle n_{\bf k} \rangle+\bar{\phi}\delta_{{\bf k},0}$, 
or

\begin{equation}
\label{defopar} 
\left \langle \phi({\bf r}) \right \rangle = \beta 
\int \rho ({\bf r}_1)  
v({\bf r}-{\bf r}_1) d {\bf r}_1 + \bar{\phi}\, 
\end{equation}
where $\rho({\bf r})=\left \langle n({\bf r}) \right \rangle$ is the average 
one-particle density. Eq.(\ref{defopar}) shows that (up to $\bar{\phi}$) 
the average order parameter at point ${\bf r}$ equals to the effective 
one-body potential (in units of $k_BT$), therefore,  
$- \langle \phi({\bf r}) \rangle $ has a similar physical meaning as 
$c^{(1)}({\bf r})$, the first-order direct correlation function in fluid 
theory \cite{Evans:92}.  
One may thus conclude  that the order parameter $\phi({\bf r})$ corresponds
to the {\it microscopic}  one-body potential within the fluid, the  
effective Hamiltonian gives energy, written in terms of the microscopic 
potential, and the partition function equals  to the (functional) integral 
over this field (see (\ref{Q3})). It is also worth to note that  
$\langle \phi({\bf r}) \rangle $  changes more smoothly in space than 
$\rho({\bf r})$, and that for homogeneous fluids 
($\rho({\bf r})=\rho={\rm const}$) 
$\langle \phi({\bf r}) \rangle = \beta v_0 \rho +\bar{\phi}={\rm const}$.

Now we calculate the surface tension using the mean field approach where 
only the extremal field $\phi^*({\bf r })$, which minimizes the free energy,  
is taken into account, i.e.  $F_{\rm mf}={\cal H}(\phi^*)$.
For the flat interface with $\nabla =d/dx$, the equation for the extremal 
field reads  \cite{Igna:2000,BrezinFeng:84}

\begin{equation}
\label{extfiel}
\kappa d^2 \phi^* / d x^2 = dV(\phi^*)/d \phi^* \, .
\end{equation} 
In the bulk of the two phases, i.e. far  from the surface, the order parameter
takes constant values, $\phi^*_1$ at $x \to -\infty$, and 
 $\phi^*_2$ at $x \to \infty$, which are related to the mean densities of 
these phases: the liquid, $\rho_l$, and the vapor, $\rho_g$ density. 
Using the fact that in the bulk 
$\phi^*_{1,2}=\beta v_0 \rho_{l,g}+\bar{\phi}$, one can show that $\phi^*_1$
and $\phi^*_2$ may be obtained by the standard double-tangent construction, 
$V^{\prime}(\phi^*_1)=V^{\prime}(\phi^*_2)$ and 
$V(\phi^*_1)+\phi^*_1 V^{\prime}(\phi^*_1)=
V^{\prime}(\phi^*_2)+\phi^*_2 V^{\prime}(\phi^*_2)$. 
If we choose interface located at $x=0$, (\ref{extfiel}) yields 
$\frac12 \kappa (d \phi^*/dx)^2=V(\phi^*)-V(\phi^*_1)$ for $x<0$ and  
$\frac12 \kappa (d \phi^*/dx)^2=V(\phi^*)-V(\phi^*_2)$ for $x>0$.
The surface tension is equal to the difference per unit area between 
the  free energy, calculated for the space-dependent $\phi^*({\bf r })$ and 
that for $\phi^*_1$ for $x<0$ and $\phi^*_2$ for $x>0$. If the 
order parameter at the interface equals $\phi^*_0$, which may be 
chosen from the condition 
$\phi^*_1<\phi^*_0< \phi^*_2$,  $V^{\prime}(\phi^*_0)=0$ the surface 
tension reads (with $V_{1,2}= V(\phi^*_{1,2})$):

\begin{equation} 
\label{surt1}
\beta \gamma = \int_{\phi^*_1}^{\phi^*_0}
\sqrt{2\kappa[V(\phi)-V_1]}d\phi
+\int_{\phi^*_0}^{\phi^*_2}
\sqrt{2\kappa[V(\phi)-V_2]}d\phi 
\end{equation} 
Now we choose the system for which the coefficient 
$h$ in (\ref{coefinal}) vanishes, i.e. $h=0$. Then for 
$V=\frac12a_2 \phi^{*2}+\frac{1}{4!}u_4\phi^{*4}$ we obtain  
$\phi^*_{1,2}=\pm (-6a_2/u_4)^{1/2}$, $\phi^*_{0}=0$, and the 
symmetric solution to Eq.(\ref{extfiel}), 
$\phi^*(x)=(-6a_2/u_4)^{1/2}\tanh(x/\xi_0)$   
($\xi_0=\sqrt{-\kappa/2a_2}$) \cite{Igna:2000,BrezinFeng:84}, with zero over 
the volume average, 
$\bar{\phi^*} \equiv \Omega^{-1}\int \phi^*({\bf r})d{\bf r}=0$.

Averaging over the volume (\ref{defopar}), yields   
$\bar{\phi^*}=\beta \bar{\rho}v_0 + \bar{\phi}$, which implies that 
$\bar{\phi}=-\beta v_0 \bar{\rho}$, with  
$\bar{\rho}=\Omega^{-1}\int \rho({\bf r})d{\bf r}= N/\Omega$ being the 
averaged over the volume density. Since $\phi^*_{1}=-\phi^*_{2}$ and 
simultaneously $\phi^*_{1,2}=\pm \beta v_0 \rho_{l,g}+\bar{\phi}$, we 
finally conclude that $\bar{\rho}= (\rho_l+\rho_g)/2$, i.e. that the 
averaged density of our 
system is the mean between the liquid and vapor density. Naturally, this 
is the density of our homogeneous reference 
system, which has the same volume and number of particles.
With the above $\phi^*_{1,2}$ and $\phi^*_{0}$,  integration 
in (\ref{surt1}) is easily performed yielding: 

\begin{equation}
\gamma/k_BT  =4 \left( 2 \kappa a_2^3/u_4^2 \right)^{1/2}
\label{surt2}
\end{equation}
where microscopic expressions for the constants  $a_2$, $u_4$, $\kappa$,
are given by Eqs.(\ref{coefinal},\ref{kappa}) in which the density 
${\rho}= (\rho_l+\rho_g)/2$ of the reference fluid is to be used. 

Not far from the critical point ($\rho_c$, $T_c$), one can 
approximate, $(\rho_l+\rho_g)/2 \simeq \rho_c$ and thus use
 $\rho_c$ as the reference density. In particular
one can write for $a_2$: 
$a_2 \simeq a_2(\beta, \rho_c)=(\beta v_0)^{-1}-\rho_c(z_0+z_3^2/2u_4)_c$ 
(see (\ref{coefinal})). If we then use the  the mean field 
condition for the critical point, $a_2(\beta_c, \rho_c)=0$ \cite{Amit:78}, 
we obtain a simple expression: 
$a_2=(\beta v_0)^{-1}-(\beta_c v_0)^{-1}=\alpha \tau$, 
and finally for the surface tension:

\begin{equation}
\gamma/k_BT = 4 \left( 2 \kappa_c \alpha^3/u_{4c}^2  \right)^{1/2}
\tau^{3/2}\, , 
\label{surt3}
\end{equation}
where $\alpha=(\beta_c v_0)^{-1}$, $\tau=(T_c-T)/T_c$, and coefficients
$u_{4,c}$ and $\kappa_c$ are to be calculated at
$\rho=\rho_c$, $T=T_c$. 

\begin{center}
\begin{figure}[tbh]
\vspace{-0.2cm}
\narrowtext
\rotatebox{0}{
\epsfxsize=8.0cm
\epsfbox{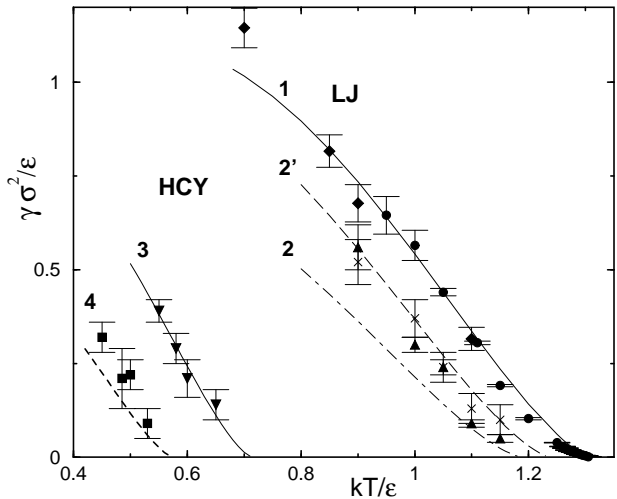}}
\caption{Reduced surface tension $\gamma \sigma^2/k_BT$ as a function 
of the reduced temperature $k_BT/\epsilon$ for the Lennard-Jones (LJ) and 
hard-core Yukawa (HCY) fluid.
Curves -- theory, Eq.(\ref{surt3}):
$1$ -- LJ-fluid; $2$, $(2')$, $3$, $4$ -- HCY-fluid for $\lambda=1.8$, 
$\lambda=3.0$ and $\lambda=4.0$ correspondingly. Points --
numerical data:
diamonds -- LJ (MD \cite{MDsur}), 
circles -- LJ (MC \cite{Panagiotopoulos:2001}),
stars and triangles up  -- HCY for $\lambda=1.8$ 
(MC and MD \cite{Trokhymchuk:2001}), 
triangles down -- HCY for $\lambda=3.0$ 
(MD \cite{Trokhymchuk:2001}), 
squares -- HCY for $\lambda=4.0$ 
(MD \cite{Trokhymchuk:2001}).   
Critical parameters are taken from \cite{Panagiotopoulos:2001} for 
the LJ-fluid, from \cite{Lomba:1994} (curves 2-4) and from 
\cite{Duh:1997} (curve $2^{\prime}$) for the HCY-fluid.  
$\sigma$, $\epsilon$ and $\sigma$, $\epsilon$, $\lambda$ are parameters 
of the LJ and HCY potential \cite{Panagiotopoulos:2001,Trokhymchuk:2001}.}
\label{fig:surten}
\end{figure}
\end{center}

The theoretical expression for the surface tension (\ref{surt3}) has been  
compared with available data of numerical experiments for the Lennard-Jones 
(LJ) and hard-core Yukawa (HCY) fluids, for which 
the standard WCA partition 
(see e.g. \cite{gray}) of the potential into attractive and repulsive parts
has been applied \cite{Brill:98}. 
As follows from Fig.1 our theory is in a  
good agreement with the numerical experiments, except for the very 
close vicinity of the critical point, where the mean field theory 
loses its accuracy. Eq.(\ref{surt3}) is quite  sensitive to the 
critical parameters $\rho_c$, $T_c$. While these are known quite accurately 
for the LJ fluid, they are estimated with much larger uncertainty for the HCY 
fluid. This is shown in Fig.1 where two theoretical curves 
($2$ and $2^{\prime}$) correspond to the same HCY fluid but with $\rho_c$, 
$T_c$, taken from different references 
($\rho_c$ and  $T_c$ differ by about $4\%$).

In conclusion, we derive an analytical microscopic expression for the 
surface tension which allows (for the first time to our knowledge) to obtain 
this quantity for the given temperature, density and interaction potential. 
This relation is in a good agreement with numerical experiments in the 
near-critical region. Our theory is based on the effective field theoretical 
Hamiltonian, which order parameter corresponds 
to the microscopic one-body potential and its average to the first-order direct 
correlation function. 
Although our  mean field result is not valid in the very 
 close vicinity of the critical point, its generalization 
to account the critical fluctuations (e.g. within the one-loop approximation 
\cite{BrezinFeng:84}) is straightforward, and will be addressed elsewhere
\cite{BrilRubi:2001}.

\end{document}